\documentclass[11pt,twoside]{article}
\usepackage{cozumel2005}
\usepackage{epsf}
\usepackage{psfig}
\usepackage{graphicx}
\usepackage{lscape}
\begin{document}
\title{Deriving stellar properties from photometry: maximizing information
       content and minimizing biases}    %%% Fill in title
\author{J. Ma\'{\i}z Apell\'aniz}   %%% Fill in author names
\affil{Space Telescope Science Institute, 3700 San Martin Drive, Baltimore, 
       MD 21218, U.S.A.}    %%% Fill in author affiliations
\affil{Space Telescope Division, European Space Agency, ESTEC, Noordwijk, 
       Netherlands}    %%% Fill in author affiliations

\begin{abstract} %%% Abstract to run on from here.
I study the importance of the accurate calibration of photometric systems in order to produce meaningful 
comparisons between the observed colors + magnitudes and model SEDs. Possible sources of errors are discussed
and two examples are analyzed. I show that well-calibrated Tycho-2 photometry is stable and precise 
enough for such comparisons. On the contrary, the available calibrations for Johnson $UBV$ photometry yield 
relative large systematic errors, which has prompted me to develop a new, more precise calibration. The 
advantages of multicolor photometry over the standard single-color + magnitude diagrams for the derivation of
physical properties of stars (elimination of 
degeneracies, inclusion of multiple parameters, avoidance of linearizing approximations, 
possibility of a more precise treatment of errors) are discussed through the use of CHORIZOS, a code developed
specifically for this purpose.
\end{abstract}

%%% MAIN BODY OF TEXT GOES HERE. CONSULT  manual_cozumel2005.tex 
%%% SECTIONS 2.3-2.6 FOR HELP WITH EQUATIONS, FIGURES,
%%% AND TABLES.

\section{A brief introduction to synthetic photometry}

$\,\!$\indent	The most common way of studying the properties of stellar populations is by comparing
measured single-color + magnitude diagrams (SCMDs) with the synthetic magnitudes derived from spectral
energy distributions (SEDs) by means of a synthetic photometry code. The formula used to compute
the magnitude for a photon-counting detector is:

\begin{equation}
m_{P} = -2.5\log_{10}\left(\frac{\int P(\lambda)f_{\lambda}(\lambda)\lambda\,d\lambda}
                                {\int P(\lambda)f_{\lambda{\rm,ref}}(\lambda)\lambda\,d\lambda}\right)
                                + {\rm ZP}_P
\label{mag1}
\end{equation}

\noindent and the equivalent formula for an energy-integrating detector is:

\begin{equation}
m_{P}^\prime = -2.5\log_{10}\left(\frac{\int P(\lambda)f_{\lambda}(\lambda)\,d\lambda}
                                {\int P(\lambda)f_{\lambda{\rm,ref}}(\lambda)\,d\lambda}\right)
                                + {\rm ZP}^\prime_P.
\label{mag2}
\end{equation}

\begin{figure}[!ht]
\centerline{\includegraphics*[width=\linewidth]{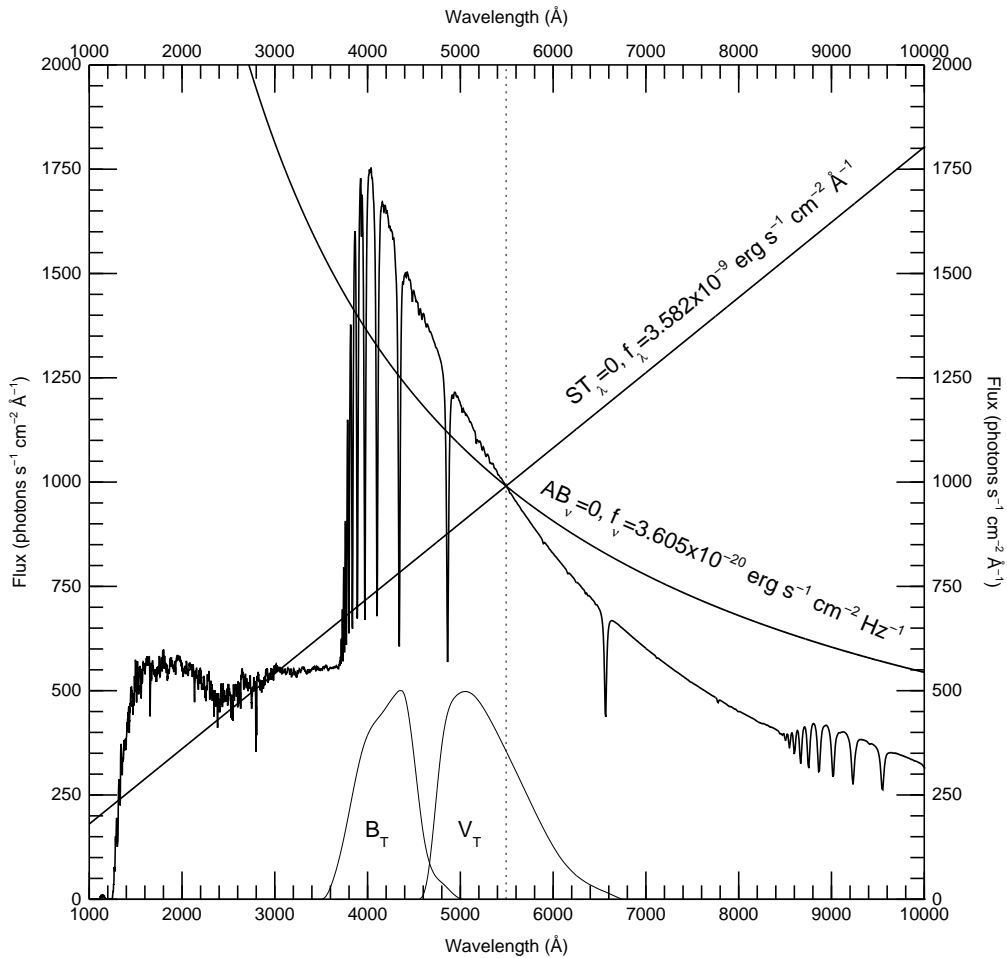}}
\caption{Reference spectra for magnitude systems: the Vega spectrum of \citet{BohlGill04}, 
the constant $f_\lambda$ spectrum for the ST system, and the constant $f_\nu$ spectrum for the
AB system. The latter two are normalized to have the same flux (in photons
s$^{-1}$ cm$^{-2}$ \AA$^{-1}$) at $\lambda = 5493$ \AA\ (dotted line). The non-normalized 
photon-counting sensitivity curves for the Tycho-2 $B_T$ and $V_T$ filters are also plotted.}
\label{refsp}
\end{figure}

	The quantities in those formulae are:

\begin{itemize}
  \item $f_\lambda(\lambda)$ is the SED of the object.
  \item $f_{\lambda{\rm,ref}}(\lambda)$ is the SED of the reference spectrum. The most common choices
	are Vega, a constant in $f_\lambda$, and a constant in $f_\nu$ (see Fig.~\ref{refsp}).
  \item $P(\lambda)$ is the total-system sensitivity curve.
  \item ZP$_P$ (or ZP$^\prime_P$) is the zero point for filter $P$.
\end{itemize}

	A magnitude system for a filter $P$ is defined by giving both $f_{\lambda{\rm,ref}}(\lambda)$ 
and ZP$_P$. Common choices are:

\begin{itemize}
  \item The VEGAMAG system uses Vega as a reference spectrum and has ZP$_P$ = 0 by definition \citep{synphot}.
  \item The STMAG system also has ZP$_P$ = 0 and as a reference spectrum uses a constant in $f_\lambda$ 
	normalized in such a way as to have the same flux as Vega at the pivot wavelength of the Johnson $V$
	filter, 5493 \AA. We obtain a value of $f_\lambda = 3.582\cdot10^{-9}$
	erg s$^{-1}$ cm$^{-2}$ \AA$^{-1}$ from the Vega spectrum of \citet{BohlGill04}.
  \item The ABMAG system also has ZP$_P$ = 0 and as a reference spectrum uses a constant in $f_\nu$ 
	normalized in such a way as to have the same flux as Vega at the pivot wavelength of the Johnson $V$
	filter, 5493 \AA. We obtain a value of $f_\nu = 3.605\cdot10^{-20}$
	erg s$^{-1}$ cm$^{-2}$ Hz$^{-1}$ from the Vega spectrum of \citet{BohlGill04}.
  \item The Johnson-Cousins system uses Vega as a reference spectrum and has values of $ZP_P$ for 
	$P=U,B,V\ldots$ that NEED TO BE MEASURED.
\end{itemize}

	There are some possible error sources that can yield biases when using synthetic photometry. The first
one is using the wrong equation (\ref{mag1} or \ref{mag2}). Since Eq.~\ref{mag1} gives more weight to longer
wavelengths, it generates brighter magnitudes for red objects than for blue ones compared to Eq.~\ref{mag2}. This can
be quantified e.g. by computing $\Delta m = (m_{P,1}-m_{P,1}^\prime)-(m_{P,2}-m_{P,2}^\prime)$ for a red (1) and a 
blue (2) SED (Table~\ref{Deltam}), since that quantity gives the maximum systematic error introduced in the analysis
of a CMD. If one wants to convert between energy-integrating and photon-counting magnitudes it is useful to
define a sensitivity curve $P^\prime(\lambda) = P(\lambda)/\lambda$. Then, it is easy to show that 
$m_{P^\prime} = m_{P}^\prime$ if ZP$_{P^\prime}$ = ZP$^\prime_P$.

\begin{table}[t]
\begin{tabular}{lccccccccc}
\hline
           & \multicolumn{5}{c}{Johnson-Cousins}   & \multicolumn{4}{c}{Str\"omgren}  \\
Filter     & $U$   & $B$   & $V$   & $R$   & $I$   & $u$   & $v$    & $b$    & $y$    \\
\hline
$\Delta m$ & 0.049 & 0.053 & 0.037 & 0.054 & 0.020 & 0.019 & 0.003  & 0.001  & 0.024  \\
\hline
\end{tabular}
\caption{$\Delta m$ for Johnson-Cousins and Str\"omgren filters using as red and blue spectra solar-metallicity
main-sequence Kurucz models with $T_{\rm eff}$ of 3\,500~K and 50\,000~K, respectively.}
\label{Deltam}
\end{table}

	A second possible source of errors is an incorrect $f_{\lambda{\rm,ref}}(\lambda)$. For the case of Vega, the
most precise spectrum is the one obtained by \citet{BohlGill04} using a combination of HST/STIS CCD spectroscopy
and Kurucz models. The absolute flux calibration has an accuracy of 4\% in the FUV and 2\% in
the optical \citep{Bohl00} but, given that the photometric repeatability of STIS is 
0.2-0.4\% \citep{Bohletal01}, the relative flux calibration for colors derived from STIS spectra 
is expected to be better than 2\% in the optical. 

	Two other possible sources of error are an incorrect knowledge of $P(\lambda)$ and of ZP$_P$. In the next two
sections we analyze the cases of the Tycho-2 $B_TV_T$ and Johnson $UBV$ filter sets.

\section{Calibration of Tycho-2 $B_TV_T$ photometry}

\begin{figure}
\centerline{\includegraphics*[width=0.62\linewidth]{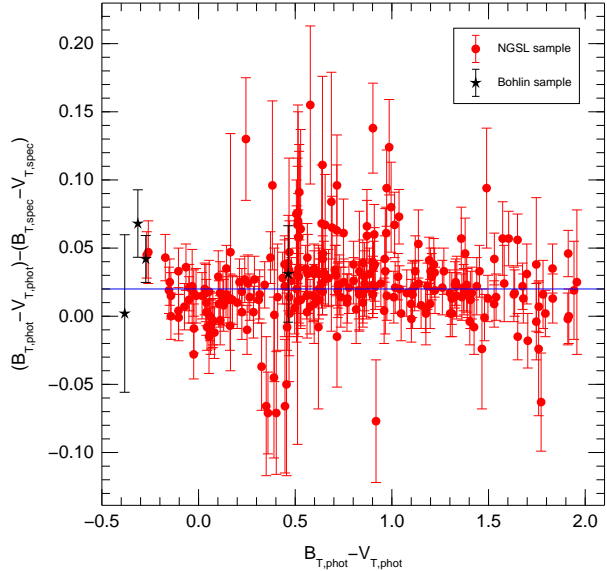}}
\caption{Comparison between photometric and spectrophotometric Tycho-2 $B_T-V_T$ colors as a function 
of photometric $B_T-V_T$ for the two samples. The error bars represent the photometric
uncertainties and the horizontal line marks the proposed ZP$_{B_T-V_T}$.}
\label{btvtplot1}
\end{figure}

$\,\!$\indent	The Hipparcos mission \citep{ESA97} observed the full sky and yielded the Tycho 
catalog, which is reasonably complete down to $V = 11.5$. The original Tycho catalog was 
consequently reprocessed by \citet{Hogetal00a} to produce the Tycho-2 catalog, which contains
2.5 million stars and is currently the most complete and accurate all-sky photometric survey in
the optical. The Tycho-2 catalog contains photometry in two optical bands, $B_T$ and $V_T$,
whose sensitivities were analyzed by \citet{Bess00} and use Vega as the reference spectrum. 
The careful processing of Tycho-2 
photometry was described by \cite{Hogetal00b}, including the different tests used to check for
possible systematic errors. 

\begin{figure}
\centerline{\includegraphics*[width=0.47\linewidth]{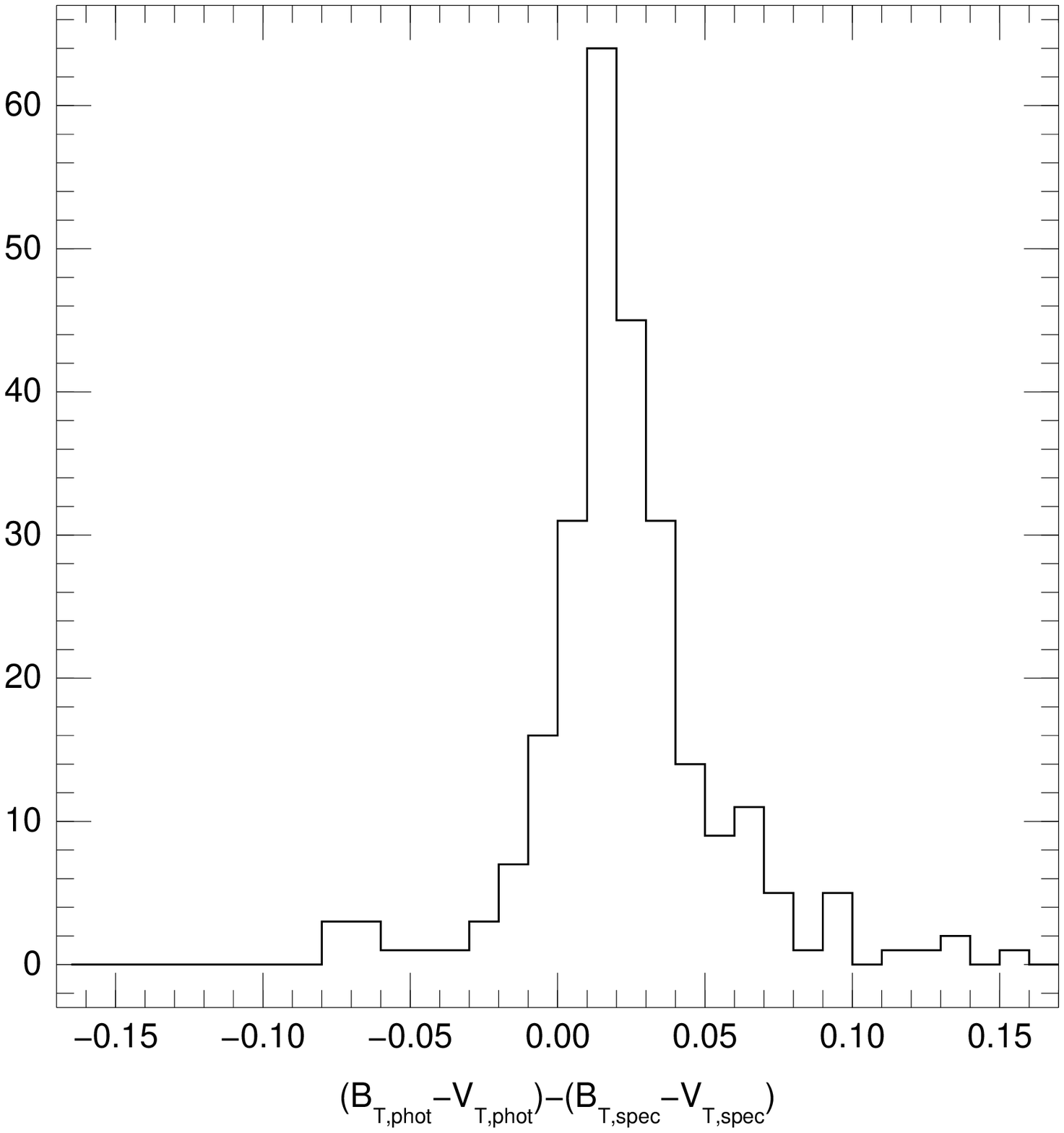}
            \includegraphics*[width=0.47\linewidth]{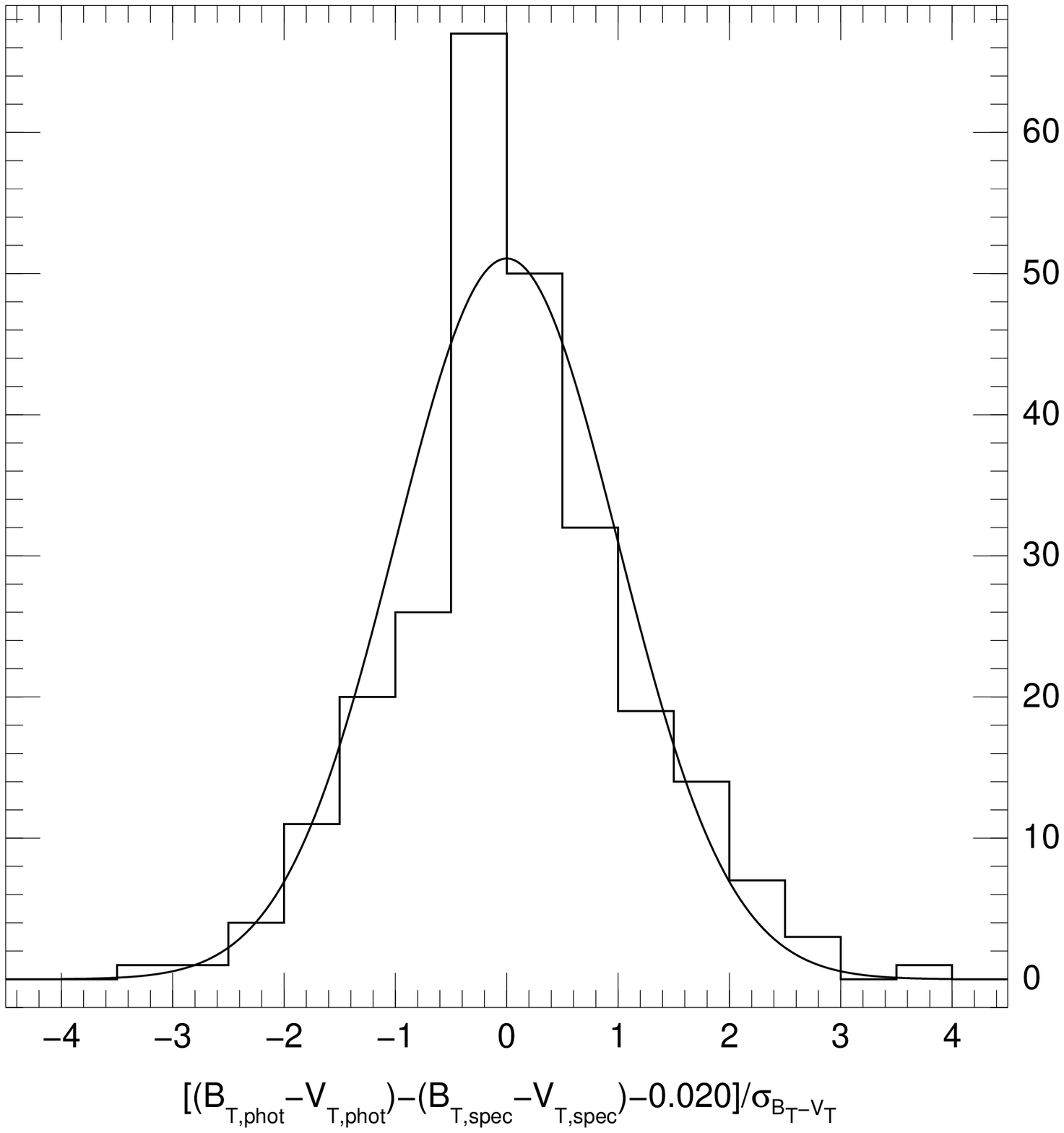}}
\caption{Histograms for the comparison between photometric and spectrophotometric $B_T-V_T$ 
colors for the NGSL sample. (left) Regular histogram. (right) Histogram for the data shifted by 
the proposed ZP$_{B_T-V_T}$ and normalized by the individual uncertainties. A Gaussian 
distribution with $\mu=0$ and $\sigma=1$ is overplotted for comparison.}
\label{btvtplot2}
\end{figure}

	I have recently \citep{Maiz05b} tested the sensitivity curves of the Tycho-2 photometry and calculated the
values of ZP$_{B_T}$, ZP$_{V_T}$, and ZP$_{B_T-V_T}$ using the spectra obtained with STIS for the 
Next Generation Spectral Library (or NGSL, 
{\tt http://lifshitz.ucdavis.edu/\~{}mgregg/gregg/ngsl/ngsl.html} and \citealt{Gregetal04}) 
and the spectrophotometric standards of
\citet{Bohletal01}. The results for $B_T-V_T$ are shown in Figs.~\ref{btvtplot1}~and~\ref{btvtplot2}. 
No general trend is observed as a function of color in Fig.~\ref{btvtplot1} and the data are symmetrically 
distributed around a central value, which I take to be
ZP$_{B_T-V_T}$. I measured ZP$_{B_T-V_T}$ by calculating the weighted mean using 
$1/\sigma^2_{B_T-V_T}$ as weights and found it to be 0.020$\pm$0.001 magnitudes.
The histograms for the $B_T-V_T$ data, both in absolute and
relative (corrected for ZP$_{B_T-V_T}$ and dividing each point by its photometric uncertainty)
terms, appear in Fig.~\ref{btvtplot2}.
The second histogram has a median of $1.2\cdot 10^{-5}$ and a standard deviation of 1.04
and the distribution is very well approximated by a normalized Gaussian. All of the above implies
that an accurate cross-calibration of colors vs. relative fluxes between Tycho-2 photometry and
HST spectrophotometry is possible in principle without having to invoke e.g. modifications in 
the Tycho filter sensitivities or the STIS calibration. Furthermore, given that the normalized
histogram has a standard deviation only slightly larger than 1.0, the largest source of
deviations from the expected value originates in the photometry, not in the spectrophotometry.
Since the mean photometric $\sigma_{B_T-V_T}$ = 0.025 magnitudes, the accuracy of the
spectrophotometrically-derived Tycho-2 colors must be better than 1\%, which agrees with the
published value for the STIS photometric repeatability \citet{Bohletal01}. A similar analysis
for the individual magnitudes $B_T$ and $V_T$ yields ZP$_{B_T}$ = 0.078$\pm$0.009 magnitudes and
ZP$_{V_T}$ = 0.058$\pm$0.009.

	These results imply that the Tycho-2 photometry has well-characterized sensitivity curves and that the
zero points are accurate and stable, both in absolute and relative terms. Therefore, it should be possible to
produce accurate comparisons between Tycho-2 photometry and synthetic
photometry generated from SED models.

\section{Calibration of Johnson $UBV$ photometry}

$\,\!$\indent	The \citet{John66} $UBV$ system is the most commonly used photometric system in the optical range. 
It was originally defined using photomultipliers but it was later adapted for CCDs. Despite its extensive use, its
calibration for a comparison with synthetic photometry has a number of problems:

\begin{itemize}
  \item As opposed to the Tycho-2 case, the Johnson photometry available in the literature has been obtained by 
	different observers using different detectors, telescopes, and observing sites, and applying different
	reduction procedures. Under those circumstances, an artificial scatter of the measured magnitudes and
	colors is inevitable.
  \item Ground-based photometry should not be as stable as space-based photometry due to the larger instability of 
	the observing conditions and the difficulties in correcting for atmospheric extinction, a problem that is
	exacerbated for broad-band photometry with respect to the intermediate- or narrow-band case due to 
	differential effects with wavelength. This point is especially important for the $U$ band, whose 
	short-wavelength limit is determined mostly by the atmosphere.
  \item It is not clear that all published synthetic photometry works have included the correct distinction between 
	Eqs.~\ref{mag1}~and~\ref{mag2} (see e.g. \citealt{synphot}).
\end{itemize}

	A consequence of the above problems can be seen in the sensitivity curves for $UBV$ published by 
\citet{BuseKuru78} and \citet{Bess90}. First, both of those articles are unable to find a unique $B$ sensitivity curve
that is capable of generating the observed $U-B$ and $B-V$ colors. Instead, they resort to two different definitions of
$B$, one for each color, a result that is clearly unphysical. Second, the $U$ sensitivity curves of those two articles
are quite different: for A-type stars, the effect on the derived synthetic magnitudes is rather small, but for
early-type O-stars the difference in the synthetic $U-B$ is almost 0.1 magnitudes. 

	In order to reduce the effect of the problems above, I am working on a recalibration of Johnson $UBV$
photometry following an approach similar to the one followed in \citet{Maiz05b} for Tycho-2 data. I have collected 
the Johnson photometry for the non-variable stars in the NGSL using the Lausanne database \citep{Mermetal97} and
selected those stars with at least four different observations in $U-B$ or $B-V$ in order to account for
different observing conditions. For the absolute calibration of the photometry, I have used the value of 
ZP$_V$ = 0.026$\pm$0.008 obtained by \citet{BohlGill04}. Here I report the preliminary results.

\begin{itemize}
  \item The measured $B-V$ colors agree well with the synthetic ones derived from the spectrophotometry and the
	sensitivity curves of either \citet{BuseKuru78} or \citet{Bess90}. ZP$_{B-V}$ is very close to zero, in 
	agreement with the result of \citet{Bessetal98}. 
  \item No combination of the \citet{BuseKuru78} and \citet{Bess90} $U$ and $B$ sensitivities is capable of 
	reproducing the observed photometry. 
  \item I have used a $\chi^2$-minimization algorithm to generate a new $U$ sensitivity curve which, when combined
	with the $B$ curve used to calculate $B-V$ colors, is capable of reproducing the observed $U-B$ colors. 
	Therefore, it is possible to eliminate the discrepancies between the measured data and the synthetic 
	photometry and, at the same time, get rid of the need of a double definition for $B$.
\end{itemize}

\section{CHORIZOS: maximizing information content}

\begin{figure}
\centerline{\includegraphics*[width=\linewidth]{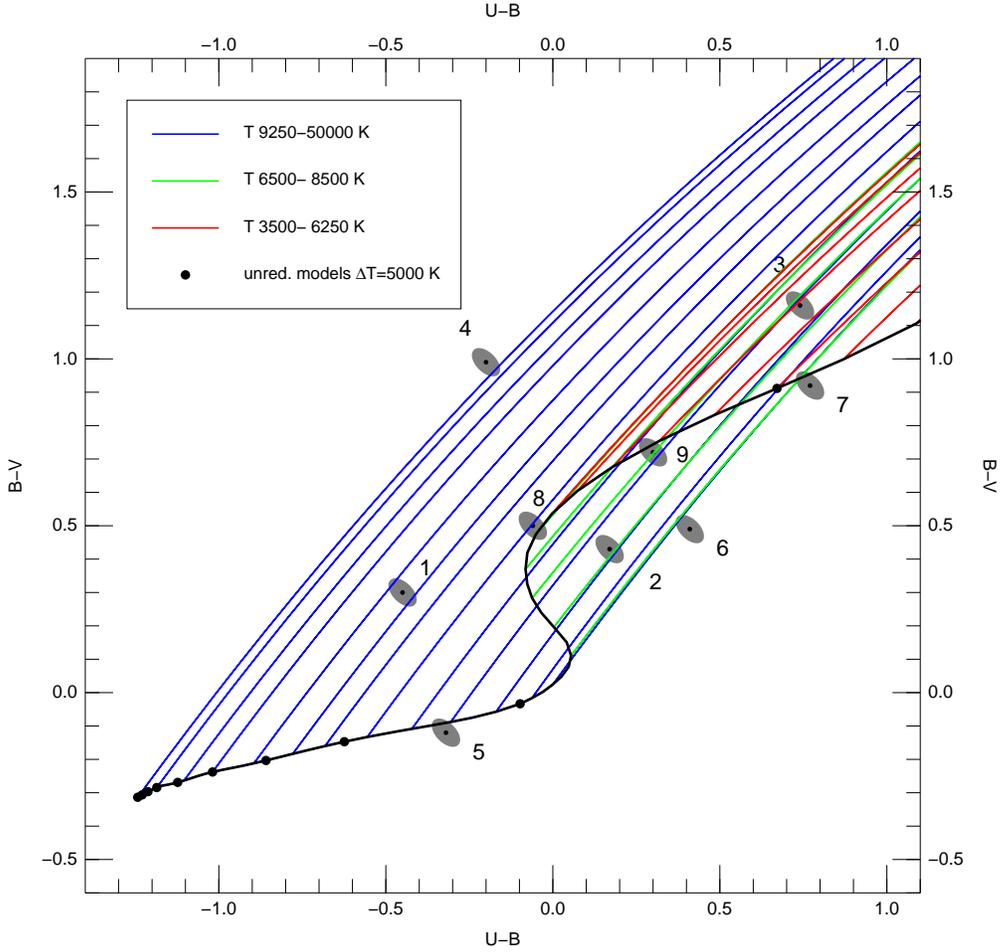}}
\caption{$U-B$ vs. $B-V$ color-color plot for $Z=0.0$, main-sequence
Kurucz atmospheric models. The line with circles indicates the location of the
unredenned values as a function of temperature, starting at $T=50\,000$ K, with
the circles marking those points where the temperature is a multiple of 5\,000 K.
The rest of the lines indicate the colors as a function of reddening using the
\citet{Cardetal89} law with $R_{5495}=3.1$, with the color code being used 
to differentiate among temperature ranges which are relevant to determine
the number of possible temperature + reddening solutions for a given $(U-B)$ 
+ $(B-V)$ combination. Nine measured stars are marked, each one of them with
$\sigma_U = \sigma_B = \sigma_V = 0.026$. Note that the reddening lines are not 
straight and that they are not parallel to each other because no Q-parameter
approximation is used.}
\label{ubvcolorcolor1}
\end{figure}

\begin{figure}
\centerline{\includegraphics*[width=\linewidth]{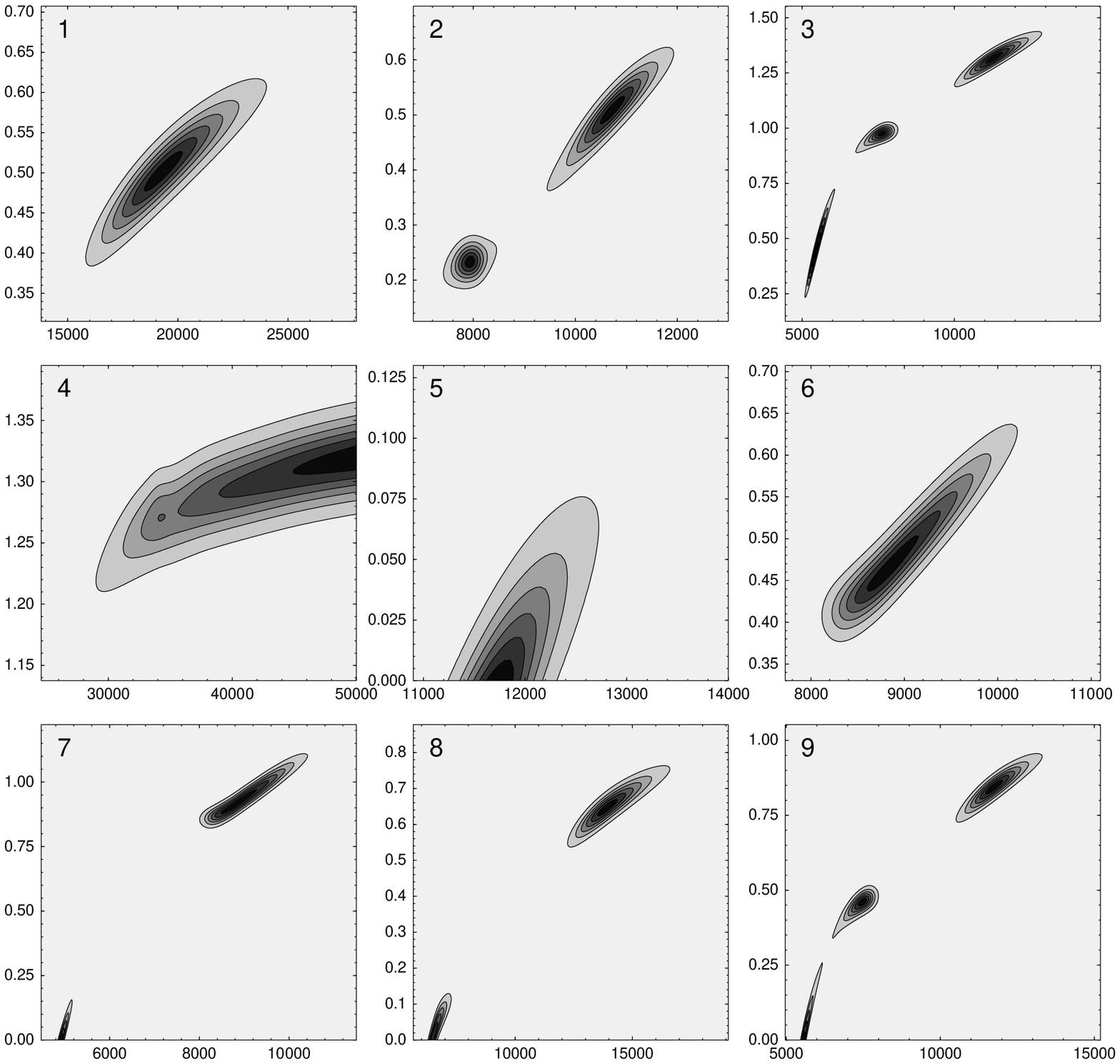}}
\caption{Likelihood contour plots produced by CHORIZOS for the nine stars
shown in Fig.~\ref{ubvcolorcolor1}. The $x$ axis
corresponds to the temperature in K and the $y$ axis to the reddening $E(4405-5495)$.}
\label{ubv1}
\end{figure}

$\,\!$\indent As previously mentioned, astronomers usually attempt to derive the $N$ parameters of individual stars 
(temperature, luminosity, metallicity, extinction\ldots) or of stellar populations
(age, metallicity, extinction\ldots) by comparing the observed SCMDs with the synthetic magnitudes derived from 
SEDs. That is, they transform their model atmosphere outputs from the theoretical (or parameter) $N-dimensional$ space
to an observational $M$-dimensional space with $M=2$ (one color and one magnitude) and compare the results with the
data there. Such a strategy has a fundamental problems: if $N>2$, degeneracies are likely to exist, so a single 
solution cannot be obtained (this can happen even if $N=2$). Also, it is not straightforward to include information from
an arbitrary number of additional colors: this is usually done by working on individual planes (e.g. projections onto
two-dimensional observational space or color-color diagrams) but in doing so one loses the full multidimensional
information. It is also common in such cases to use linearizing approximations, such as Q-ratio extinction corrections
or filter transformations, hence introducing additional systematic errors (see Fig.~\ref{ubvcolorcolor1}).

\begin{figure}
\centerline{\includegraphics*[width=0.55\linewidth]{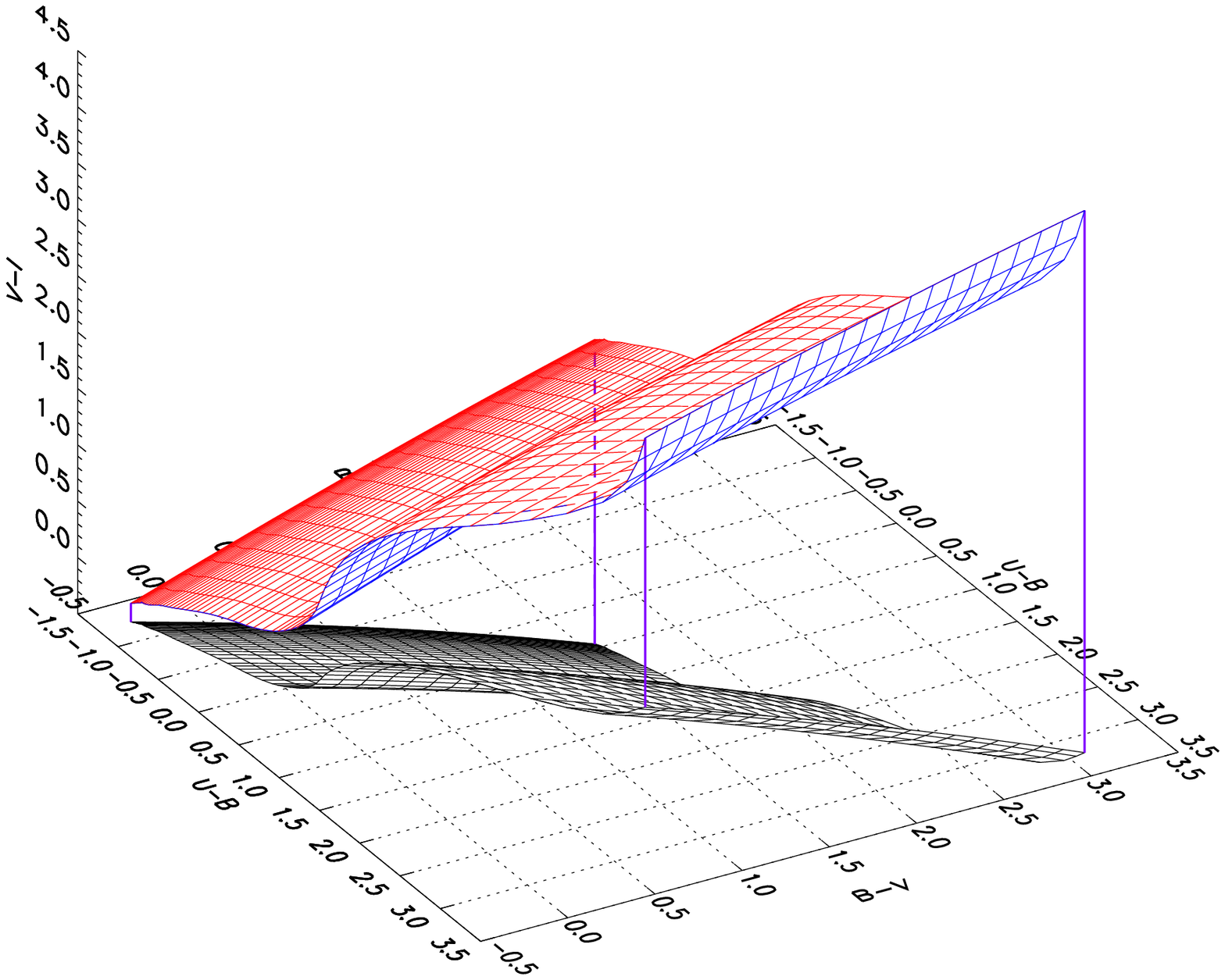}
          \ \includegraphics*[width=0.26\linewidth,bb=28 -200 566 566]{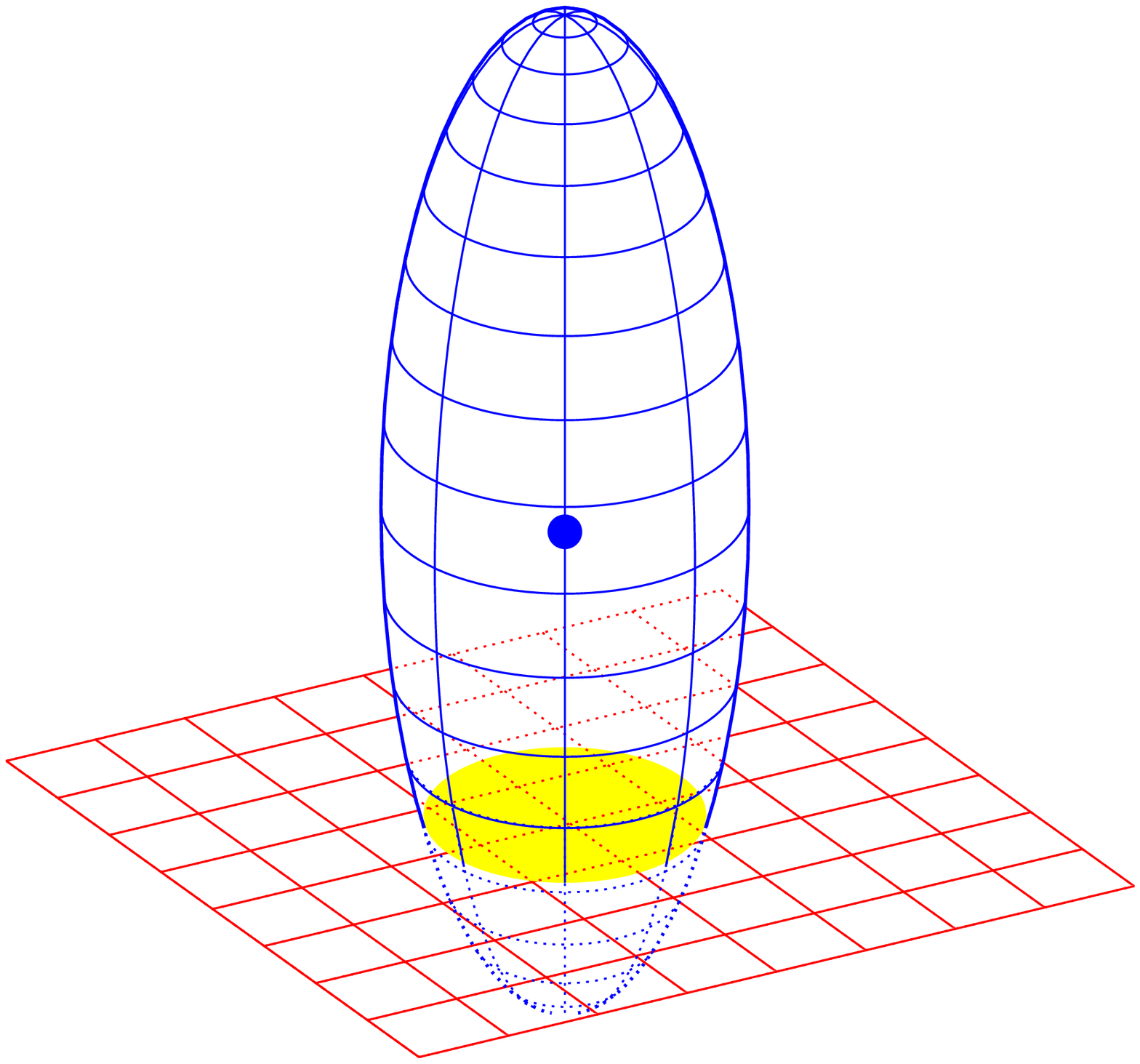}}
\caption{(left) $V-I$ vs. $B-V$ vs. $U-B$ 3-color plot for $Z=0.0$, main-sequence
Kurucz atmospheric models. The range plotted covers  $T=3\,500-50\,000$ and 
$E(4405-5495)=0.0-2.0$
and the extinction law used is that of \citet{Cardetal89} with $R_{5495}=3.1$.
The color surface marks the location in 3-color space while the black one is the projection
onto the $(U-B)$-$(B-V)$ plane (see Fig.~\ref{ubvcolorcolor1}).
(right) Basic topology for an $M=3$, $N=2$ case such as the one on the left panel. Given that 
$M>N$, the measured colors
(blue circle) always lies outside the solution surface (red grid). However, including the
uncertainty ellipsoid (blue grid), yields an intersection surface (solid yellow) of likely 
solutions.}
\label{fig3d}
\end{figure}

	An alternative is to use a code such as CHORIZOS \citep{Maiz04c}. CHORIZOS first calculates the theoretical 
colors for the full $N$-dimensional parameter space and then evaluates the likelihood in each point of the grid
for the observed photometry. The moments of
the resulting likelihood are then calculated to derive the range of possible parameters.
Furthermore, since the code does not operate by simply finding a likelihood maximum (or $\chi^2$ minimum) it allows for
the detection of multiple solutions. A simple example for the well-known case of finding the temperature and 
extinction ($N=2$) for stars of known metallicity and gravity and with a standard extinction law from Johnson 
$U-B$, $B-V$ colors ($M=2$) is shown in 
Figs.~\ref{ubvcolorcolor1}~and~\ref{ubv1}. Such an example has degeneracies, as it can be seen in the two possible
solutions for e.g. star 2 and in the three possible solutions for e.g. star 3. Such degeneracies can be lifted by using
additional colors (Fig.~\ref{fig3d}), which CHORIZOS can easily handle.

	CHORIZOS is available to the astronomical community and can be downloaded from 
{\tt http://www.stsci.edu/\~{}jmaiz}. A number of upgrades 
to the code have been added since the publication of the original article \citep{Maiz04c} and more are planned for the
near future. We detail the current and future capabilities here:

\noindent\begin{tabular}{p{0.468\linewidth}p{0.468\linewidth}}
 & \\
\multicolumn{1}{c}{CURRENT} & \multicolumn{1}{c}{FUTURE} \\
 & \\
{\bf SEDs:} Kurucz, Lejeune, TLUSTY (stars); Starburst99 (clusters). &
Arbitrary user-defined SED models for stars, clusters, galaxies\ldots \\
 & \\
{\bf Parameters:} Up to four. Two SED-intrinsic ($T_{\rm eff}$ + log$g$, age) and two extrinsic
(reddening + extinction law). &
Five, user-defined and in any combination (metallicity, redshift, IMF slope, upper mass limit\ldots). \\
 & \\
{\bf Parameter control:} Full range, restricted range, fixed. Grid size adjustable and extendable. &
Use of Bayesian priors \\
 & \\
{\bf Filters:} 80 preinstalled (Johnson-Cousins, Str\"omgren, Tycho-2, SDSS, 2MASS, HST). &
Future HST instruments?, user-defined. \\
 & \\
{\bf Spectrophotometry:} No. & Yes. \\
 & \\
\end{tabular}

	Eventually, CHORIZOS will become a full Bayesian code that will be able to handle any arbitrary SED
family or filter set.

\begin{figure}
\includegraphics*[width=0.476\linewidth,bb=32  52.0 542 530.0]{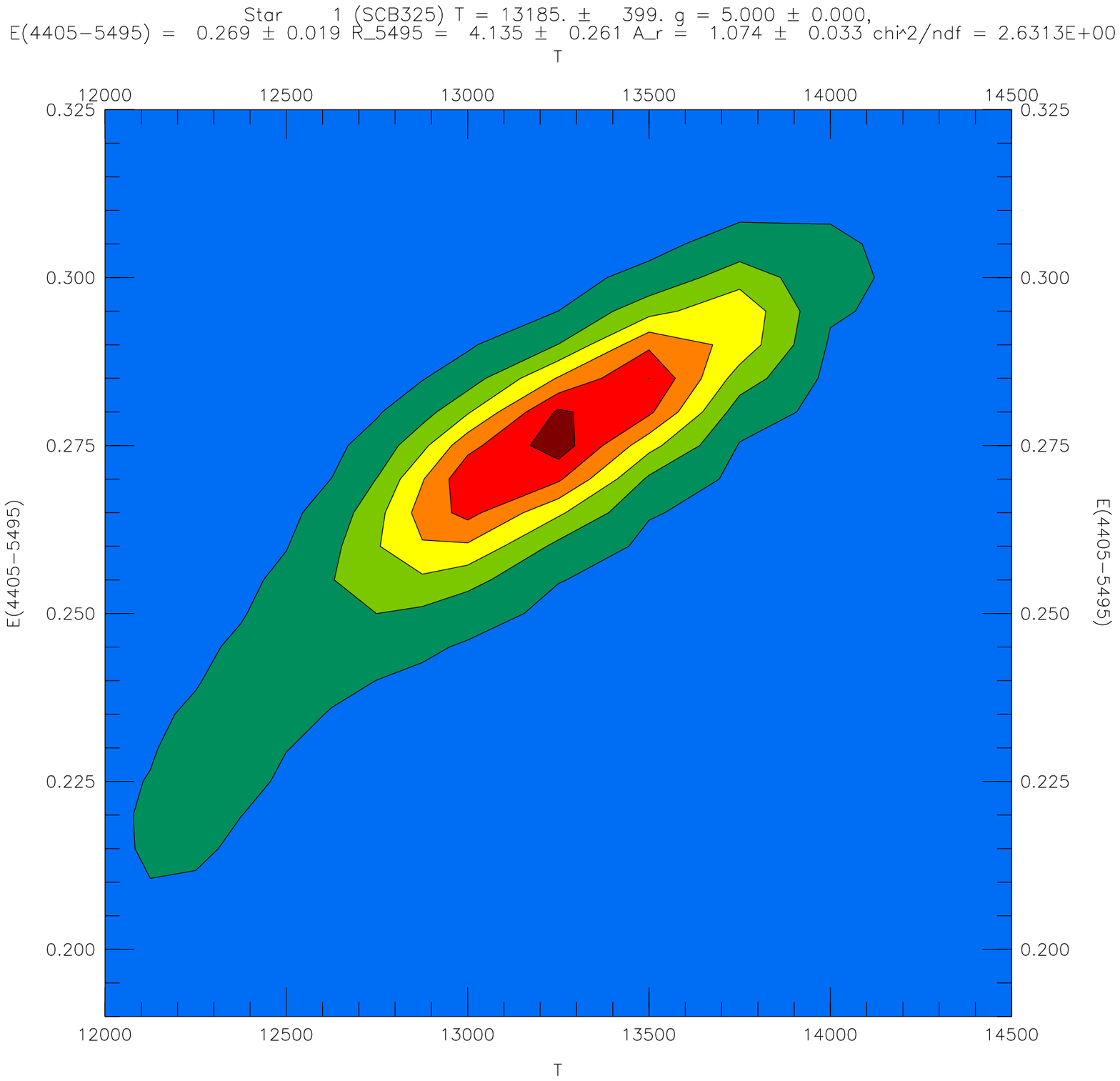} \\ \
\includegraphics*[width=0.470\linewidth,bb=28  28.0 542 520.0]{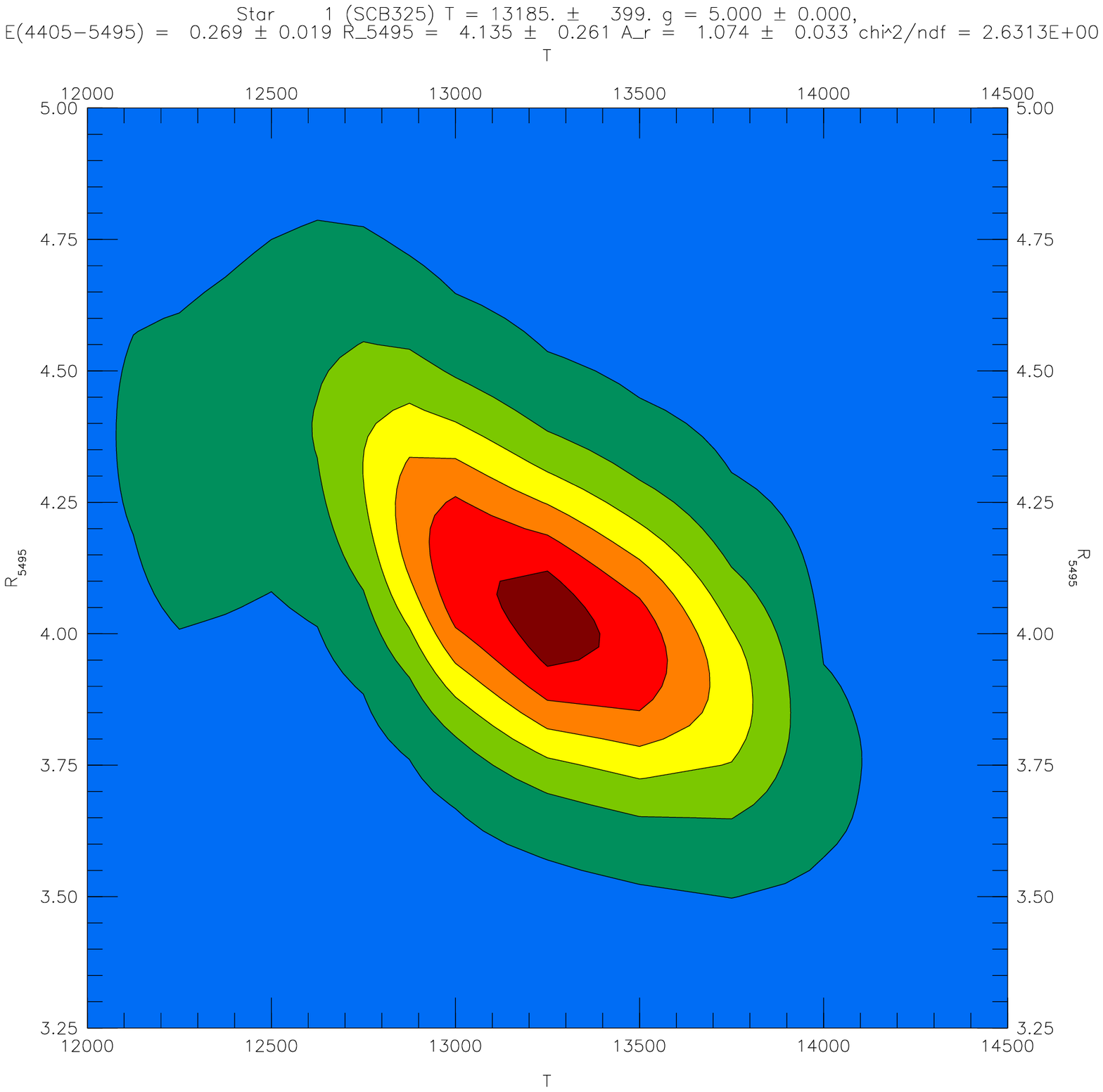} \
\includegraphics*[width=0.470\linewidth,bb=51  26.5 567 518.5]{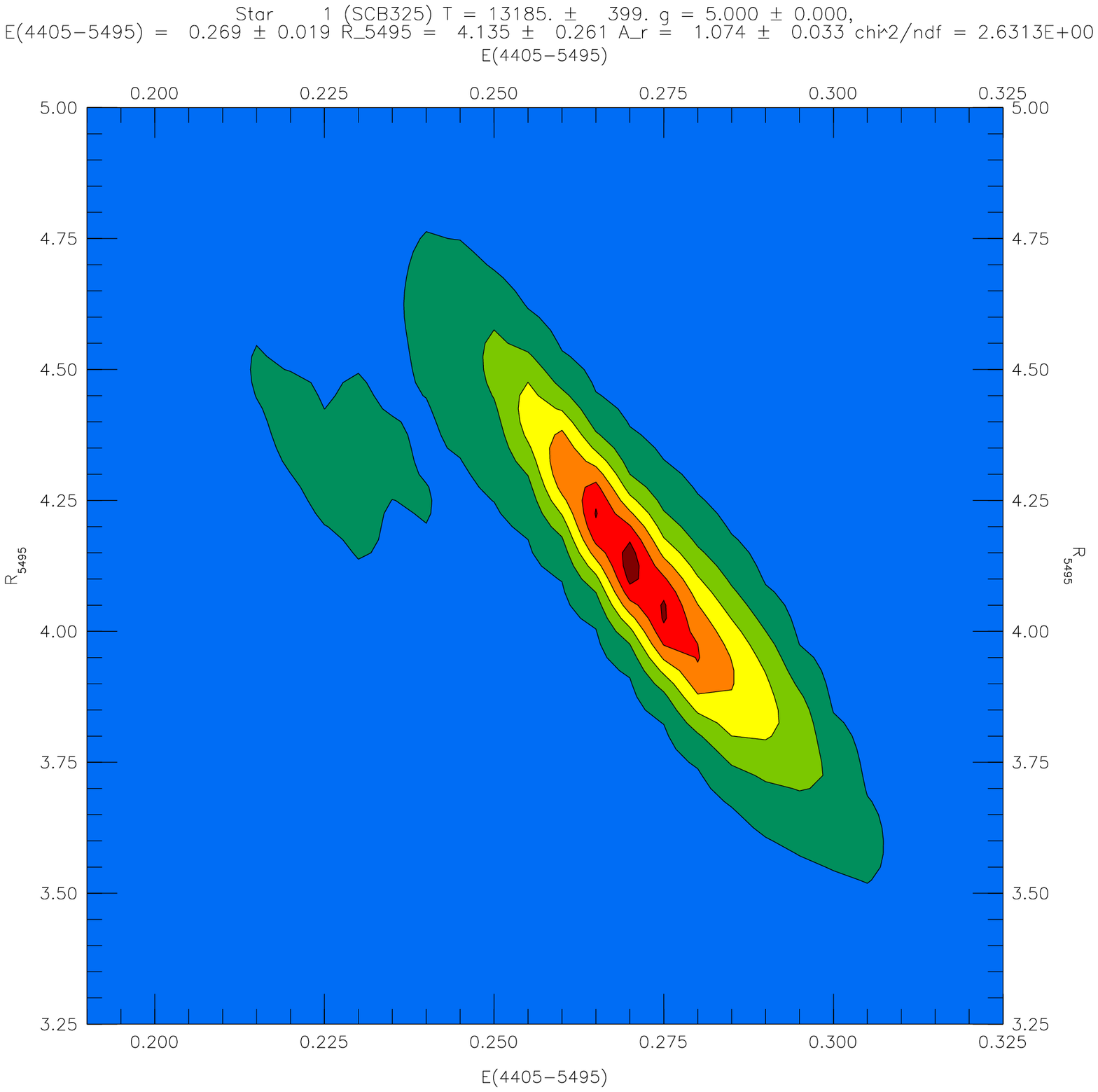} 
\caption{Likelihood contour plots produced by CHORIZOS for one of the stars in the M8 sample. 
Each plot corresponds to the projection onto one of the 3 orthogonal planes defined using
the three coordinates $T_{\rm eff}$, $E(4405-5495)$, and $R_{5495}$: 
$T_{\rm eff} - E(4405-5495)$ (upper left), $T_{\rm eff} - R_{5495}$ (lower left), and
$E(4405-5495) - R_{5495}$ (lower right).}
\label{m8_1}
\end{figure}

	As an example of the possible applications of CHORIZOS to the study of stellar populations we include here
a summary of the study performed by \citet{Ariaetal05}. Those authors analyzed $UBVIJHK_{\rm s}$ photometry for six
stars in the Galactic H\.{\sc ii} region M8. CHORIZOS was executed using solar-metallicity, main-sequence Kurucz
stellar models with three free parameters: temperature, reddening, and extinction law. CHORIZOS parameterizes
reddening by $E(4405-5495)$, the monochromatic equivalent of $E(B-V)$ and extinction law by $R_{5495}$, the 
monochromatic equivalent to $R_V$. The extinction laws of \citet{Cardetal89} were used. 

	The likelihood plots for SCB 325, one of the six stars is shown in Fig.~\ref{m8_1}. 
CHORIZOS produces as graphical output the
projection of the likelihood into each of the possible combinations of two free parameters (in this case, there are
three free parameters, yielding three possible combinations). As it can be seen, the output is a well-defined peak with
only a slight asymmetry, so the values of $T_{\rm eff}$, $E(4405-5495)$, and $R_{5495}$ and their uncertainties can be
derived with confidence.

\begin{figure}
\centerline{\includegraphics*[width=0.48\linewidth]{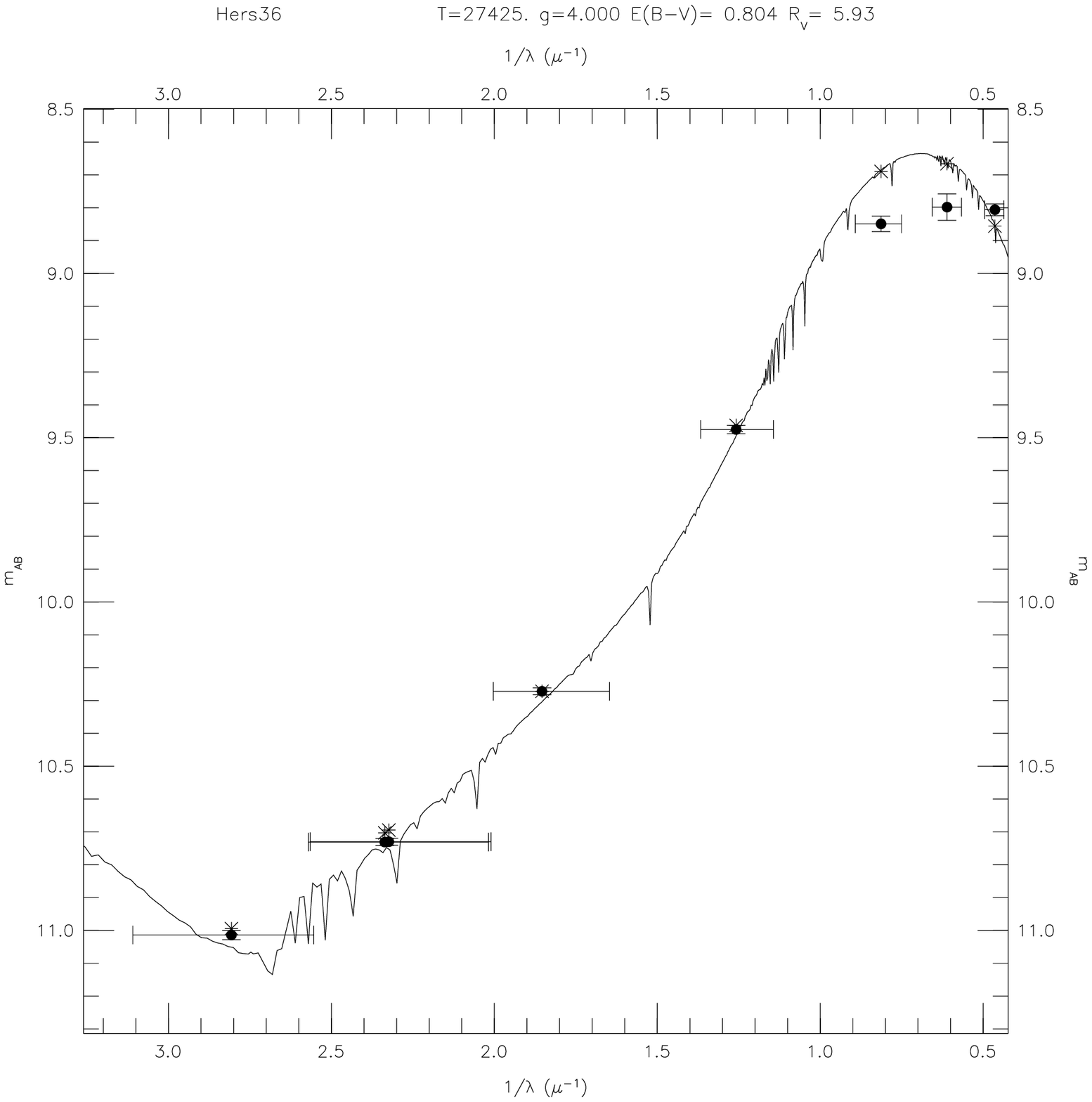} \
            \includegraphics*[width=0.48\linewidth]{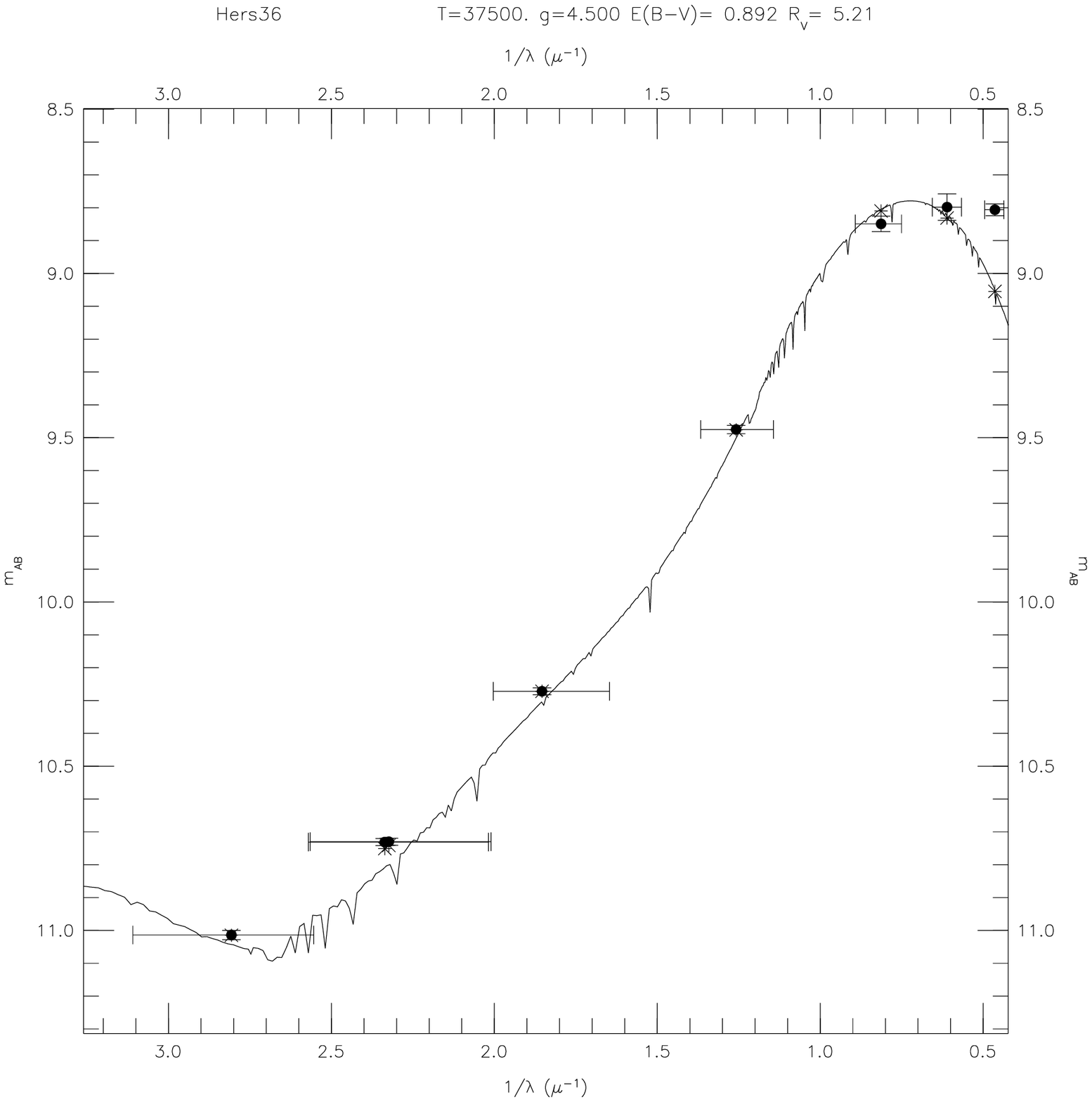}}
\caption{Measured photometry and best-fit spectrum for Herschel 36 produced by CHORIZOS. The left plot shows the
fit using all seven photometric bands ($UBVIJHK_{\rm s}$) while the right plot uses only the
first five ($UBVIJ$). The corresponding values of the reduced minimum $\chi^2$ are 21 and
1.7, respectively. For the photometry, the vertical error bars are used to show the photometric uncertainties
while the horizontal ones show the approximate extent of each filter in wavelength.}
\label{m8_2}
\end{figure}

	Another of the CHORIZOS outputs is shown in Fig.~\ref{m8_2} for the case of Herschel 36, another star in the
M8 sample. The left plot shows the best-fit spectrum when all seven photometric bands (six colors) are used. Such a
best fit has a reduced $\chi^2$ of 21, indicating that the observed photometry is not compatible with the SED 
family and/or the parameter range used. A look at the plot immediately suggests the origin of the problem: the $UBVI$
photometry is apparently well fitted but the NIR data is not. As it turns out, Herschel 36 has a large IR excess, so a
Kurucz model is not a good approximation for its SED in the NIR: CHORIZOS detects such circumstance by yielding a 
reduced minimum $\chi^2 \gg 1$. If on the other hand, CHORIZOS is run excluding the $HK_{\rm s}$ photometry (right
plot of Fig.~\ref{m8_2}), the result is very different: the reduced $\chi^2$ for the best fit is close to 1, as it is
readily apparent from the concordance between the spectrum and the photometry (note that the two rightmost photometric
points are not included in the fit). 

\begin{figure}
\centerline{\includegraphics*[width=\linewidth]{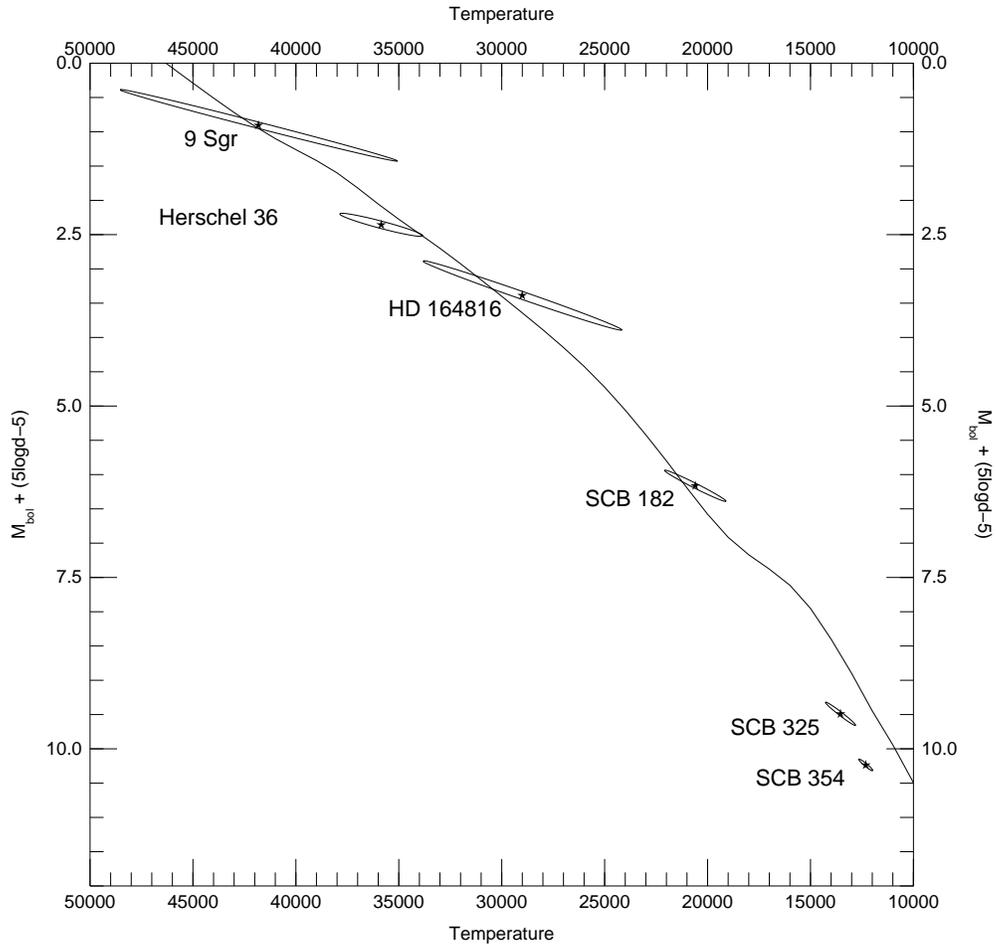}}
\caption{Theoretical ( $T_{\rm eff}$ vs. $M_{\rm bol} + 5\log d - 5$) HR diagram for the 
six stars in the M8 sample produced by CHORIZOS. Ellipses indicate the 68\% likelihood
contour for each star. The line shows the main sequence for $5\log d - 5 = 10.5$. As seen
in the plot, such a distance is compatible with the CHORIZOS output for the four earliest 
stars but not for the latest two.}
\label{m8_3}
\end{figure}

	Once the photometry of each star has been processed by CHORIZOS and the parameters 
($T_{\rm eff}$, $E(4405-5495)$, $R_{5495}$) and their uncertainties have been evaluated, it is possible to 
construct the theoretical HR diagram (Fig.~\ref{m8_3}). Note that the uncertainty ellipses are inclined with 
respect to the two coordinates. The reason is the strong correlation between $T_{\rm eff}$ and $M_{\rm bol}$
induced by the dependence of the bolometric correction on temperature. The HR diagram shows that the 
photometry of the four earliest-type stars in the sample is compatible with a distance modulus of 10.5. The two late B
stars, however, must be farther away, since their uncertainty ellipses fall clearly below the expected main sequence at
that distance. 

\section{Conclusions}

$\,\!$\indent	I have shown the importance of the accurate calibration of photometric systems in
order to produce meaningful comparisons between the observed colors and magnitudes and model
SEDs. The Tycho-2 calibration derived by \citet{Bess00} (for the sensitivity curves) and 
\citet{Maiz05b} (for the zero points) is precise enough for that task. On the other hand, the available 
calibrations for Johnson $UBV$ photometry yield relatively large systematic errors, which has prompted
me to develop a new, more precise calibration. I have also shown how the use of multicolor photometry 
has significant advantages over the standard single-color + magnitude diagrams, among them the 
elimination of degeneracies, the inclusion of multiple parameters, the avoidance of linearizing
approximations, and the possibility of a precise treatment of errors.

\acknowledgements             %%% Text of acknowledgements runs on after this command.

I would like to thank Ralph Bohlin and Rodolfo Barb\'a for fruitful conversations about the topics 
discussed in this work.

%%% THE BIBLIOGRAPHY
%%%
%%% CONSULT SECTION 3 OF   manual_cozumel2005.tex    FOR HOW TO USE NATBIB.
%%% AUTHORS ARE ENCOURAGED TO USE EITHER THE "THEBIBLIOGRAPY" ENVIRONMENT
%%% or THE BIBTEX ENVIRONMENT. 

\bibliographystyle{aj}
\bibliography{general}

%\begin{thebibliography}{}
%\bibitem[Bland-Hawthorn \& Jones(1998a)]{bh1998a} Bland-Hawthorn, J., \& Jones, D. H. 1998a, \apj, 15, 44
%\end{thebibliography}

\end{document}